\documentstyle[twoside,fleqn,espcrc2]{article}
\newcommand{\be}{\begin{equation}}
\newcommand{\ee}{\end{equation}}
\newcommand{\bea}{\begin{eqnarray}}
\newcommand{\eea}{\end{eqnarray}}

\newcommand{\no}{\nonumber\\}

\newcommand{\tAp}{\widetilde{{\mathbf{A}}}_p}
\newcommand{\tCp}{\widetilde{{\mathbf{C}}}_p}

\newcommand{\bO}{{\mathbf{O}}}
\newcommand{\cS}{{\cal S}}
\newcommand{\cC}{{\cal C}}

\newcommand{\rmga}{(r-\gamma_{\alpha})}

\newcommand{\NC}{N_c}
\newcommand{\NS}{N}

\newcommand{\AmS}{{\protect\the\textfont2
  A\kern-.1667em\lower.5ex\hbox{M}\kern-.125emS}}

\title{$\NS=1$ Super Yang-Mills on the Lattice in the Strong Coupling Limit}

\author{E. Gabrielli, A. Gonz\'alez-Arroyo and C. Pena
\address{Departamento de  F\'{\i}sica Te\'orica,
Universidad Aut\'onoma de Madrid,
28049 Madrid, Spain}
}
       
\begin{document}

\begin{abstract}
We study the  $\NS=1$ supersymmetric $SU(\NC)$ Yang-Mills theory
on the lattice at strong coupling. 
We analyse and discuss the recent results obtained at strong coupling
and large $\NC$ for the mesonic and fermionic propagators and 
spectrum.
\end{abstract}

\maketitle

\section{Introduction}
The pure $\NS=1$ SUSY Yang-Mills (SYM) theory with 
$SU(\NC)$ gauge group is described by the purely gluonic action plus
one flavour of Majorana fermions in the adjoint representation of the 
colour group.

The non-perturbative aspects of this class of theories 
were intensely investigated in the past\cite{npSUSY},
and recently there has been a renewed interest on the
subject \cite{shifman}.
On the other hand, these theories can be studied on the lattice by using the standard
lattice methods.
However, when  supersymmetric gauge theories are formulated 
on the lattice, some problems arise due to the fact that the
lattice regularization spoils supersymmetry. Nevertheless, in Ref.~\cite{cv}
it is shown how to recover supersymmetry in the continuum limit
by using appropriately renormalized operators for the
SUSY and chiral currents.

Two different collaborations have carried out
numerical analysis of the SYM on the lattice
\cite{andrea,montvay} (quenched and unquenched respectively)
by following the guidelines of Ref.~\cite{cv}.
The main reason for performing an unquenched analysis from the beginning
is that in the SYM theory the quenched approximation cannot be justified
(in the continuum limit) on the basis of large $\NC$ dominance. 
However, in Ref.~\cite{andrea} it is shown that the results on the spectrum, 
obtained by employing the quenched approximation, match the supersymmetry 
predictions within their  statistical errors.

Our work aims at complementing these numerical studies with analytical
information on the strong coupling, large $\NC$ region of the theory. 
While not being  the physical (i.e. continuum) regime, this limit allows 
to extract relevant information on the dynamics of the lattice theory, and 
supplies a guideline for Monte Carlo simulations.

 Leaving supersymmetry aside, our  results can be regarded as an 
analysis of   the spectrum  of Yang-Mills fields coupled to Majorana quarks 
in the adjoint representation of the group. The large $\NC$, strong 
coupling analysis  reported in Refs.~\cite{our,largeN} was performed in 
arbitrary space-time dimension. In the present paper, we summarize the 
main results of Ref.~\cite{our}.

\section{Strong Coupling}

We begin by writing down the lattice  action we will be working with:
\be
S = \beta S_g + \frac{1}{2} \Psi_i \Psi_j {\bf M}_{i j }\ \ ,  
\ee
where $\beta S_g$ is the pure gauge part and $\Psi_i$ is a Grassman variable
representing  the field of a Majorana fermion. 
Note that the indices $i$ and $j$ run over space-time,  colour and Dirac 
indices.
The matrix ${\bf M}$ must be
antisymmetric and its form (in compact notation) is given by
\bea
{\bf M}_{i j} = C\left[{\mathbf{I}} - \kappa \sum_{\alpha} \delta_{m\, n+\alpha}\
U_{\alpha}(n)\,(r {\mathbf{I}} - \gamma_{\alpha})\right]
\eea
where $\mathbf{I}$ and $C$ are the unit and charge conjugation matrices respectively,
$r$ is the Wilson parameter,
$\kappa$ is the hopping parameter and  $n,m$ parametrize  lattice points.
The  index $\alpha$ labels the $2d$ possible nearest neighbour steps and their
corresponding lattice vectors. For forward steps
$U_{\alpha}(n)$ labels  the adjoint gauge link variable and $\gamma_{\alpha}$
the Dirac matrices.

The lattice strong coupling expansion is an analytical method
which provides information on the behaviour of the lattice gauge theories
at small values of $\beta$ and $\kappa$ (hopping parameter).
Our precise technique can be summarized as follows:
\begin{itemize}
\item As customarily done, we replace the  Pfaffian ${\mbox{Pf}}({\bf M})$
(which appears due to the Majorana character of the fermions) by the positive 
square root of the determinant. This is indeed rigorous for 
$|\kappa| < \frac{1}{2 d\, (|r|+1)}$. 
\item
The hopping parameter expansion around the origin allows to 
 express  the propagator $\mathbf{M^{-1}}$ and the determinant 
${\mbox{det}}({\bf M})$  as sums over paths on the lattice.
\item The integration over the group is performed at large $\NC$.
The only leading lattice paths are those with zero area (pure spike paths 
\cite{largeN}).
\item The leading lattice path resummation is performed 
for any $r$ and in any dimension \cite{combi}.
In our analysis we keep ${r}$ arbitrary 
because this allows for the possibility of searching for 
multicritical points.
\end{itemize}

A first crucial result is that, by taking into account the 
$SU(\NC)$ group integration at large $\NC$~ 
\cite{largeN},
the quenched and OZI approximations
turn out to be exact in the large $\NC$ limit.

\section{Results on propagators and spectrum}
We concentrate upon gauge invariant operators
of the form:
\be
\bO_i(x)=\Psi_{A_1}^{a_1}(x) \ldots \Psi_{A_p}^{a_p}(x)\,
(\cS_i)_{A_1 \ldots A_p}\, \cC_i^{a_1 \ldots a_p} ~,
\label{pgluino}
\ee
where $\lbrace \cS_i \rbrace$ and $\lbrace \cC_i \rbrace$ are complete sets 
of spin and invariant colour tensors, respectively, of rank $p$. Note that when $p=2$ a basis
for $\cS_i$ is provided by the Clifford algebra basis in $d$ dimensions.
\\
Our specific aim is to compute the following quantities at strong coupling
\be
\langle\bO_i(x)\rangle,~~~~
G_{i j }(x-y) \equiv \langle\bO_i(x) \bO_j(y)\rangle \ \ ,
\nonumber
\ee
where as usual the $\langle~\rangle$ means vacuum expectation value.
We compute the main formulas
(which are valid at large $\NC$ and in any dimension $d$) 
for the propagators of the $p$-gluino operators
$G_{i j}(x)$  obtained after resumming over paths. The main result for
propagators is:
\bea
&&G_{i j}(x) =R_p(\xi)\,
\prod_{\mu} (\int \frac{d\varphi_{\mu}}{2 \pi})\, e^{\imath \varphi \cdot x}
\no
&&
\times 
\langle \cS_i | 
\, \lbrack \Theta_p(\xi){\mathbf{I}} - \tAp(\varphi)
\rbrack^{-1} \tCp^{-1}| \cS_j \rangle \ \ ,
\label{propag}
\eea
where $\mathbf{I}$ is the unit matrix and
\bea
\tCp \, &\equiv& \, 
\underbrace{C \otimes C \cdots \otimes C}_p\no
\widetilde{\mathbf{A}}_p(\varphi) &\equiv& \kappa^p\, 
\sum_{\alpha \in I}  e^{\imath
\varphi_{\alpha} }
\underbrace {\rmga \otimes \ldots \otimes \rmga}_p \,\no
\Theta_p(x) &\equiv& (1-x)^p + \frac{x^p}{(2d-1)^{p-1}}\no 
\xi &\equiv& \frac{1-\sqrt{1-4(2d-1)\kappa^2(r^2-1)}}{2} \ \ ,
\eea
and where $C$ is the charge conjugation matrix.
The reader is referred to Ref.~\cite{our} for 
the expressions of the functions $R_{2,3}(x)$, as well as
for the definition of the kets $|\cS_i\rangle$.

In order to obtain analytical formulas for the propagator 
and masses, one needs to
invert the matrix $\Theta_p(\xi){\mathbf{I}}-\tAp(\varphi)$ appearing
in Eq.(\ref{propag}). This can be done in an elegant way 
for arbitrary space-time dimension 
and $p=2$  by mapping the 
problem into a finite dimensional many-body 
fermion problem ({\it gamma-fermions}  \cite{our}). 

The results
for the scalar mass $M_S$ and 
the lightest pseudoscalar mass $M_P$, which would enter the
lowest supermultiplet according to the analysis of Veneziano and Yankielowicz
~\cite{npSUSY},
are, for any even dimension, the following:
\bea
\cosh(M_{S})&=&|\Phi_2(\xi) \pm (d-1)|\no
\cosh(M_{P})&=&\theta\, \Xi - H \no
H &=&\sqrt{(\theta^2 -1)(\Xi^2-1)
+ (d-1)^2 \epsilon^2}\no
\Phi_2(x)&=&\frac{(1-x)^2(2d-1)+x^2}{2x(1-x)} \no
\Xi&=&\Phi_2(\xi)-(d-1)r^2\epsilon~,
\eea
where the $\pm$ in $\cosh(M_S)$ is for $|r|<1$ and $|r|>1$ respectively, and
$\epsilon\equiv 1/(r^2-1)$, $\theta\equiv \epsilon (r^2+1)$.

The corresponding critical line (where the lightest meson becomes
massless) marks the edge of the physical region as well as 
the boundary of validity of our formulae. The equation for this critical
line is
\be
\Phi_2(\xi)=d\theta \ \ .
\ee

In odd dimensions (where the pseudoscalar operator does not exist)
the lightest state in the 2-gluino sector 
is a vectorial state and its corresponding mass is given by
\be
\cosh(M_{V})=|\Phi_2(\xi) -\theta(d-1)|~.\no
\ee

The obtention of general expressions for the spectrum valid in arbitrary 
dimension  for $p>2$ is  in general a complicated task. We 
were able to do it only in some special cases. For example, for  $p=3$
with a completely antisymmetric spin tensor one gets  a spin $1/2$ state
 whose mass is given by:
\bea
\cosh(M_{\frac{1}{2}})&=& \ \vline \, r \Xi_3 \pm \sqrt{ (\Xi_3)^2-\epsilon} \ 
\vline \no
\Xi_3 &=& \frac{\Theta_3(\xi) \epsilon^2}{2 \kappa^3} - r \epsilon
\sigma,\no
\Theta_3(x) &\equiv& (1-x)^3 + \frac{x^3}{(2d-1)^2},\no
\sigma&=&\sum_i \cos(\varphi_i=0,\pi)~.
\eea
Finally, we found that, in the $(\kappa,r)$ plane, there is a special limit where 
all the states in each $p$-gluino sector become degenerate.
This limit is obtained by letting 
$r\to \infty$ and $\kappa\to 0$ in such a way that the product $\kappa r$ is fixed.
In this limit the spin-independent $p$-gluino mass is given by:
\bea
\cosh(M_p)&=& 
\Phi\left(\left(\frac{(2d-1)(1-\xi)}{\xi}\right)^{p/2}\right)  -\sigma,\no
\Phi(x) &=& \frac{1}{2} \left( x + \frac{(2d-1)}{x} \right)~.
\eea
Masses increase with $p$.

\section{Conclusions}
The main results of our work are: 
\begin{itemize}
\item  As mentioned earlier, the quenched and OZI approximations
are exact in this limit. 
\item The pseudoscalar meson is the lightest meson for even space-time
dimensions. For odd dimensions the lightest state is the vector meson.
The vanishing of their lattice masses marks the critical line. 
\item The only multicritical point where several meson masses vanish 
corresponds to the point $\kappa \rightarrow 0$, $r  \rightarrow \infty$
with $\kappa r = \frac{1}{2 \sqrt{2d-1}} $.
\item For the  3 and 4-dimensional cases, there are no multicritical lines
with vanishing 3-gluino masses. The masses of the 3-gluino fermionic states
are always positive within the physical region of the $(\kappa,r)$ plane.
\end{itemize}

\section*{Acknowledgments}
E.G. acknowledges the financial support of the TMR network project 
ref. FMRX-CT96-0090.

\end{document}